\documentclass[times,preprint]{elsarticle}

\makeatletter
\def\ps@pprintTitle{%
 \let\@oddhead\@empty
 \let\@evenhead\@empty
 \def\@oddfoot{}%
 \let\@evenfoot\@oddfoot}
\makeatother

\usepackage{graphicx}
\usepackage{xfrac}
\usepackage{amsmath}
\usepackage{mathrsfs}
\usepackage{amsfonts}
\usepackage{lineno}
\usepackage{hyphenat}
\usepackage{cases}
\usepackage {mathtools}
\usepackage {bm}
\usepackage{amsthm}
\usepackage{amssymb}
\usepackage{lineno}
\usepackage{cuted}
\usepackage{flushend}
\usepackage[english]{babel}
\usepackage{lipsum}
\usepackage{cleveref}
\usepackage{float}
\usepackage{nicefrac}
\usepackage{textcomp}

\newtheorem{remark}{remark}

\journal{}

\begin{document}

\begin{frontmatter}

\title{Optimal Control of Spins by Analytical Lie Algebraic Derivatives}

\author[1]{Mohammadali Foroozandeh\corref{M. Foroozandeh}}
\address[1]{Chemistry Research Laboratory, University of Oxford, Mansfield Road, Oxford OX1 3TA, UK}
\cortext[M. Foroozandeh]{Corresponding author}
\ead{mohammadali.foroozandeh@chem.ox.ac.uk}

\author[2]{Pranav Singh\corref{P. Singh}}
\address[2]{Department of Mathematical Sciences, University of Bath, Bath BA2 7AY, UK}
\cortext[P. Singh]{Corresponding author}
\ead{ps2106@bath.ac.uk}

\begin{abstract}
Computation of derivatives (gradient and Hessian) of a fidelity function is one of the most crucial steps in many optimization algorithms. Having access to accurate methods to calculate these derivatives is even more desired where the optimization process requires propagation of these calculations over many steps, which is in particular important in optimal control of spin systems. Here we propose a novel numerical approach, ESCALADE (Efficient Spin Control using Analytical Lie Algebraic Derivatives) that offers the exact first and second derivatives of the fidelity function by taking advantage of the properties of the Lie group of $2\times 2$ Hermitian matrices, $\mathrm{SU}(2)$, and its Lie algebra, the Lie algebra of skew-Hermitian matrices, $\mathfrak{su}(2)$. A full mathematical treatment of the proposed method along with some numerical examples are presented.
\end{abstract}


\end{frontmatter}

\section{Introduction}
\label{sec:intro}
Controlling quantum spin dynamics using time-dependent Hamiltonians in the form of pulses (e.g. radiofrequency, microwave, and laser pulses) is the essence of method development in many areas of science \cite{RN133}, from magnetic resonance spectroscopy and imaging \cite{RN144} and terahertz technologies\cite{RN229,RN228} to trapped ions \cite{RN226}, cold atoms \cite{RN227} and NV-centers in diamond \cite{RN232,RN231} for quantum information processing and computing \cite{RN236}.

In NMR and ESR, in particular, designing radiofrequency and microwave pulses for robust excitation of signals over a very wide range of frequencies and reduced sensitivity to instrumental imperfections is still among the most challenging areas of method design and is of great interest. The development of methods for pulse design in these applications generally follows one or more of three distinct routes: composite pulse design \cite{RN65,RN87,RN67,RN105,RN63,RN88}, evolutionary numerical methods like optimal control theory (OCT) \cite{RN142,RN145,RN144,RN147,RN184,RN146,RN152,RN185}, and design of swept-frequency pulses \cite{RN139,RN11,RN12,RN207,RN37,RN34,RN54,RN8,RN81,RN190,RN218}.

Two of the main challenges in the field of optimal control of spin systems are the controllability of the dynamics and the convergence rate of the control process. In principle, three main approaches can be considered when optimal control has been applied to spin systems: 1) derivative-free techniques \cite{RN230,RN233} which are especially important when due to experimental requirements not many iterations or function evaluations by the optimisation protocol can be allowed, 2) gradient-based techniques like GRAPE \cite{RN144,RN236} and KROTOV \cite{RN235,RN234}, and 3) Newton--Raphson method \cite{RN124,RN135} where in addition to the gradient (first derivative), the Hessian (second derivative of the objective function with respect to the control parameters) is also utilized. Although the latter approach results in quadratic convergence rate, it suffers from numerical complexity due to computation and update of a dense Hession matrix in the course of optimization. Additionally, computation of derivatives using finite differences can be expensive, inaccurate and potentially unstable when the objective function involves numerical propagators with limited accuracy \cite[Chapter~8]{NoceWrig06}. Therefore having access to the exact form of these derivatives is of great interest, in particular in the optimal control of spin systems, where the optimization process requires propagation of these calculations over many steps, and inaccurate estimations of derivatives can result in a large accumulated numerical error.

The objective of this paper is to present a novel approach that facilitates the optimal control of spins using Newton--Raphson utilizing an analytical computation of derivatives. Control of dynamical systems using properties of Lie groups and their algebras covers a surprisingly wide range of applications from controlling of landing a plane, rotations of rigid bodies in robotics and estimation of camera poses in computer vision, to the time evolution of quantum systems \cite{RN224,RN220,RN219,RN221,RN222,RN225,RN223,Drummond2000,Subbarao2005}. In these applications, the underlying geometric structure is described by a Lie group. Finite difference methods suffer from a further disadvantage here since they do not respect Lie group structure and result in derivatives that do not live in the tangent space (the Lie algebra) \cite{Taylor94}.

Here we propose a novel numerical approach, ESCALADE (Efficient Spin Control using Analytical Lie Algebraic Derivatives) that harnesses the exact first and second derivatives of the fidelity function. These derivatives are computed by exploiting the properties of the Lie group of $2\times 2$ Hermitian matrices, $\mathrm{SU}(2)$, and its Lie algebra -- the Lie algebra of skew-Hermitian matrices, $\mathfrak{su}(2)$. Since the Lie groups, $\mathrm{SU}(2)$ and $\mathrm{SO}(3)$ are closely related (see \cite[Chapter~5]{Kosmann2010} and \cite[Chapter~6]{Woit2017}), there is a close parallel between some of the properties exploited here to the Rodrigues rotation formula \cite{Rodrigues1840}, which is utilized widely in computer vision and robotics applications for computation of rotation matrices in $\mathrm{SO}(3)$ \cite{Taylor94,Subbarao2005,Gallego2015,Terzakis2018}.

Although here we present the technique on the optimal control of the dynamic of non-interacting qubits, this is a general approach and can be applied to spin systems with more diverse Hamiltonian structures. It has the potential to find applications in a variety of areas where taking advantage of Lie algebra for efficient optimal control of spins is beneficial. Examples include geometric \cite{RN216,RN217,RN224,RN238,RN237} and adiabatic optimal control \cite{RN139,RN122,RN164,RN215} methods.

\section{Theory}

\subsection{Optimal control of spin-\sfrac{1}{2}}
The state of a single spin-\sfrac{1}{2} particle is described by the density
matrix $\rho(t) \in \mathrm{SU}(2)$ and its dynamics are governed by the
Liouville--von Neumann equation,
\begin{equation}
\label{eq:LvN}
\partial_t \rho(t) = -\mathrm{i} [ \mathcal{H}(t), \rho(t)], \qquad \rho(0) = \rho_0 \in \mathrm{SU}(2),
\end{equation}
where
\begin{equation}
\mathcal{H}(t)=\boldsymbol{h}(t) \cdot \boldsymbol{\sigma},
\end{equation}
\begin{equation}
\boldsymbol{h}(t) = (f(t),g(t),\Omega)^\top \in \mathbb{R}^3, \quad \boldsymbol{\sigma} = (\sigma_x, \sigma_y, \sigma_z)^\top,
\end{equation}
and
\begin{equation*}
\sigma_{x}=\frac{1}{2}\left(\begin{array}{cc}
{0} & {1} \\
{1} & {0}
\end{array}\right), \ \ \sigma_{y}=\frac{1}{2}\left(\begin{array}{cc}
{0} & {-\mathrm{i}} \\
{\mathrm{i}} & {0}
\end{array}\right), \ \ \sigma_{z}=\frac{1}{2}\left(\begin{array}{cc}
{1} & {0} \\
{0} & {-1}
\end{array}\right)
\end{equation*}
are the normalized Pauli matrices. $\Omega$ describes the offset
frequency of a spin.

\begin{remark}
In magnetic resonance applications, it is
typical to write $f(t) = \omega(t) \cos (\phi(t))$ and $g(t) = \omega(t) \sin
(\phi(t))$ where the amplitude $\omega(t)$ and the phase $\phi(t)$ may be
arbitrary (real-valued) functions of time.
\end{remark}

\begin{remark}
In the case of multiple non-interacting spin-\sfrac{1}{2} particles, the
$k$th spin evolves under the influence of $\mathcal{H}_k(t) = \boldsymbol{h}_k(t) \cdot
\boldsymbol{\sigma}$, where the offset $\Omega_k$ in  $\boldsymbol{h}_k(t) =
(f(t),g(t),\Omega_k)$ varies with the particle but $f(t)$ and $g(t)$ are
common across all spins.
\end{remark}

%
%the density matrix $\rho(t) = \rho_1(t) \otimes
%    \ldots \otimes \rho_M(t)$ remains uncoupled if the initial

%The vector space of $2 \times 2$ skew-Hermitian matrices, $\GG{su}(2)$, is
%the Lie algebra of $\mathrm{SU}(2)$, the Lie group of $2 \times 2$ unitary
%matrices. Moreover, the Pauli matrices scaled by $\mathrm{i}$ form the basis of
%$\GG{su}(2)$. Consequently all $2 \times 2$ unitary matrices can be described
%in the form $\exp(-\mathrm{i} \MM{s} \cdot \MM{\sigma})$ for some $\MM{s} \in
%\BB{R}^3$.

In a numerical solution of equation~(\ref{eq:LvN}), we compute $\rho$ at time intervals
$t_0, t_1, \ldots t_N$, with the unitary numerical propagation being
described by
\begin{equation}
\label{eq:Upropstep}
\rho_n = \mathrm{U}_{n} \rho_{n-1} \mathrm{U}_{n}^{\dagger}, \qquad \mathrm{U}_{n} = \mathrm{e}^{-\mathrm{i} \boldsymbol{s}_n \cdot \boldsymbol{\sigma}},\end{equation}
where
\begin{equation}
\label{eq:s}
\boldsymbol{s}_n =
(\Delta t)\,  (f(t_{n-1}),g(t_{n-1}), \Omega)^\top.
\end{equation}

By using equation~(\ref{eq:Upropstep}), one can see that the final density matrix is
given by
\begin{equation}
\label{eq:Uprop}
\rho_N = \mathrm{U}_{\mathrm{tot}} \rho_0 \mathrm{U}_{\mathrm{tot}}^{\dagger}
\end{equation}
where
\begin{equation}
\label{eq:Utot} \mathrm{U}_{\mathrm{tot}} = \mathrm{U}_N \mathrm{U}_{N-1} \ldots \mathrm{U}_2 \mathrm{U}_{1}.
\end{equation}

Typically we want to maximize the fidelity functional,
\begin{linenomath*}
\[ \mathcal{F} = \langle \varrho | \rho_N \rangle := \mathrm{Tr}(\varrho^{\dagger}
\rho_N ) \quad \in [0,1], \]
\end{linenomath*}
to have maximum overlap (i.e. $\mathcal{F} = 1$)
with the (normalized) target state $\varrho \in \mathrm{SU}(2)$.

In a gradient-based optimization scheme one  needs to
compute the gradient of the fidelity function $\mathcal{F}$,
\begin{equation}
\label{eq:gradF}
\frac{\partial
\mathcal{F}}{\partial \theta_{n,k}} = \mathrm{Tr}\left(\varrho^{\dagger} \frac{\partial
\rho_N}{\partial \theta_{n,k}} \right),
\end{equation}
where $n \in \{1, \ldots, N\}$, $k \in \{1,2\}$, and
\begin{linenomath*}
\[ \theta_{n,1} = f(t_{n-1}), \qquad \theta_{n,2} = g(t_{n-1})\]
\end{linenomath*}
are the control parameters that solely affect the $n$th propagator, $\mathrm{U}_n$. A Newton--Raphson optimization scheme also requires the Hessian,
\begin{equation}
\label{eq:hessF} \frac{\partial^2
\mathcal{F}}{\partial \theta_{m,j} \partial \theta_{n,k}} = \mathrm{Tr}\left(\varrho^{\dagger}
\frac{\partial^2 \rho_N}{\partial \theta_{m,j} \partial \theta_{n,k}} \right),
 \end{equation}
where $n,m \in \{1, \ldots, N\}$ and $j,k \in \{1,2\}$.

In the computation of the gradient of the fidelity function (\ref{eq:gradF}), we require the gradient of the final state $\rho_N$,
\begin{linenomath*}
\begin{equation}
\label{eq:gradrhoN}
 \frac{\partial \rho_N}{\partial \theta_{n,k}}  = 2 \mathrm{Re} \left(\frac{\partial \mathrm{U}_{\mathrm{tot}}}{\partial \theta_{n,k}} \rho_0 \mathrm{U}_{\mathrm{tot}}^{\dagger}\right).
\end{equation}
\end{linenomath*}
Since $\theta_{n,k}$ only affects the $n$th propagator, this gradient can be written in the form
\begin{linenomath*}
\begin{align}
\nonumber \frac{\partial \mathrm{U}_{\mathrm{tot}}}{\partial \theta_{n,k}} & =  \mathrm{U}_{N} \mathrm{U}_{N-1} \ldots \mathrm{U}_{n+1} \frac{\partial \mathrm{U}_n}{\partial \theta_{n,k}} \mathrm{U}_{n-1} \ldots \mathrm{U}_{1}\\
\label{eq:gradLR}& =  \mathrm{L}_{n+1} \frac{\partial \mathrm{U}_n}{\partial \theta_{n,k}} \mathrm{R}_{n-1},
\end{align}
\end{linenomath*}
where
\begin{linenomath*}
\begin{align}
\label{eq:L}\mathrm{L}_n & = \mathrm{U}_{N} \mathrm{U}_{N-1} \ldots \mathrm{U}_{n},\\
\label{eq:R} \mathrm{R}_n & = \mathrm{U}_{n} \mathrm{U}_{n-1} \ldots \mathrm{U}_{1},
\end{align}
\end{linenomath*}
can be computed in $\mathcal{O}(N)$ time.

Here we present a method for computing the gradient $\partial \mathrm{U}_n/\partial \theta_{n,k}$, and therefore the gradient of the fidelity function, analytically using Lie algebraic techniques. This approach is also extended for computing the Hessian analytically.

%with respect to a vector of control parameters  $\MM{\theta} = (\theta_1,
%\theta_2, \ldots, \theta_K)$. These parameters affect the components $f(t)$
%and $g(t)$, and consequently the evolution of $\rho(t)$. For instance, in the
%case of a standard GRAPE algorithm \cite{RN144} with piecewise constant
%pulses $f$ and $g$, we have $K = 2N$ parameters, given by the piecewise
%constant values of $f$ and $g$,
%\[\theta_{2j} = f(t_{j-1}), \qquad \theta_{2j+1} = g(t_{j-1}), \qquad \ j = 1, \ldots,
%N.\]

% $\MM{\theta} = (f(t_0), f(t_1), \ldots
%f(t_{N-1}), g(t_0), g(t_1), \ldots g(t_{N-1}))$ and $K = 2N$.

%The gradient of the fidelity functional is given by
%
%
%\vspace{2cm}
%
%. Which is where we need $\frac{\partial \rho}{\partial \theta_j}$ where
%$\theta_i,\ i = 1,\ldots, K$ is a control parameter that affects $f$ and $g$,
%and therefore the propagation of $\rho$.

%(remark on regularization and penalty functions - intro or conclusions etc).

\label{subsec:OCT}

\subsection{Computation of gradient}
In this section we present the analytic approach for computing the derivative of the $n$th unitary propagator (\ref{eq:Upropstep}),
\begin{linenomath*}
\[\mathrm{U}_{n} = \exp(-\mathrm{i} \boldsymbol{s}_n(\theta_{n,k})), \]
\end{linenomath*}
with respect to a control parameter $\theta_{n,k}$. Here we write $\boldsymbol{s}_n(\theta_{n,k})$ to highlight the fact that $\boldsymbol{s}_n$ depends on $\theta_{n,k}$.

In general, the derivative of the exponential of $X(\theta)$ with respect to a control parameter $\theta$ can be expressed as \cite{Schur1891,Rossmann2006}
\begin{equation}
\label{eq:dexp1}
\frac{\partial }{\partial \theta} \exp(X(\theta))= \exp(X(\theta))\, \mathrm{dexp}_{X(\theta)} X'(\theta),
\end{equation}
where the dexp function,
\begin{linenomath*}
\begin{equation}
\label{eq:dexp2}
\mathrm{dexp}_X X' = \left( \frac{1- \mathrm{e}^{-\mathrm{ad}_X}}{\mathrm{ad}_X} \right) (X') = \sum_{p=0}^\infty \frac{(-1)^p}{(p+1)!} \mathrm{ad}_X^p (X'),
\end{equation}
\end{linenomath*}
is expressed as a power series of the adjoint operator, $\mathrm{ad}$. The powers of $\mathrm{ad}$ are given by
\begin{align*}
\mathrm{ad}_X^0 (X') &= X',\\
\mathrm{ad}_X(X') &= [X,X'], \\
\mathrm{ad}_X^2(X') &= [X,[X,X']].
\end{align*}
Equations~(\ref{eq:dexp1}) and (\ref{eq:dexp2}) allow us to express the derivative of $\mathrm{U}_n$,
\begin{linenomath*}
\begin{align*}
\frac{\partial \mathrm{U}_n}{\partial \theta_{n,k}} &=  \frac{\partial }{\partial \theta_{n,k}} \mathrm{e}^{-\mathrm{i} \boldsymbol{s}_n(\theta_{n,k}) \cdot \boldsymbol{\sigma}}\\
& = \underbrace{\mathrm{e}^{-\mathrm{i} \boldsymbol{s}_n \cdot \boldsymbol{\sigma}}}_{\mathrm{U}_n} \left(
\frac{1 - \mathrm{e}^{-\mathrm{ad}_{-\mathrm{i} \boldsymbol{s}_n \cdot \boldsymbol{\sigma}}}}{\mathrm{ad}_{-\mathrm{i} \boldsymbol{s}_n \cdot \boldsymbol{\sigma}}}
\right)
\left( - \mathrm{i} \frac{\partial \boldsymbol{s}_n }{\partial \theta_{n,k}} \cdot \boldsymbol{\sigma} \right)\\
&=  \mathrm{U}_n \left(\sum_{p=0}^\infty \frac{(-1)^p }{(p+1)!} \mathrm{ad}^p_{-\mathrm{i} \boldsymbol{s}_n \cdot \boldsymbol{\sigma}}
\right)
\left( - \mathrm{i} \frac{\partial \boldsymbol{s}_n }{\partial \theta_{n,k}} \cdot \boldsymbol{\sigma} \right).
\end{align*}
\end{linenomath*}
\begin{remark}
For ease of notation, we suppress the dependence of $\boldsymbol{s}$ on the control parameters, $\theta$.
\end{remark}

An explicit formula can be derived for the $\mathrm{dexp}$ series when $X(\theta) \in \mathrm{su}(2)$. To see this, we introduce the map $\sim$ which maps vectors in $\mathbb{R}^3$ to matrices in $\mathrm{su}(2)$,
\begin{linenomath*}
\[ \widetilde{\boldsymbol{s}} = - \mathrm{i} \boldsymbol{s} \cdot \boldsymbol{\sigma}, \qquad \boldsymbol{s} \in \mathbb{R}^3. \]
\end{linenomath*}
%This map is closely related to the `hat' map \cite[Section~1.1.4]{Leok2018} .
It is easy to verify that
\begin{equation}
\label{eq:adhat}
 [\widetilde{\boldsymbol{s}}, \widetilde{\boldsymbol{r}}] = \widetilde{\boldsymbol{s} \times \boldsymbol{r}}= \widetilde{\boldsymbol{S} \boldsymbol{r}}, \qquad \boldsymbol{s},\boldsymbol{r} \in \mathbb{R}^3,
\end{equation}
where $\times$ is the cross product and $\boldsymbol{S}$ is the matrix,
\begin{equation}
\label{eq:S}  \boldsymbol{S} = \left(
      \begin{array}{ccc}
        0 & -s_z & s_y \\
        s_z & 0 & -s_x \\
        -s_y & s_x & 0 \\
      \end{array}
    \right).
\end{equation}

%\begin{align*}
%[\boldsymbol{s} \cdot \boldsymbol{\sigma}, \boldsymbol{r} \cdot \boldsymbol{\sigma}] & =
%[s_x \sigma_x, r_y \sigma_y] + [s_x \sigma_x, r_z \sigma_z]\\
%& \quad +[s_y \sigma_y, r_x \sigma_x] + [s_y \sigma_y, r_z \sigma_z]\\
%& \quad +[s_z \sigma_z, r_x \sigma_x] + [s_z \sigma_z, r_y \sigma_y]\\
%&= \mathrm{i} (s_y r_z - s_z r_y) \sigma_x + \mathrm{i} (s_z r_x - s_x r_z) \sigma_y + \mathrm{i} (s_x r_y - s_y r_x) \sigma_z\\
%&= \mathrm{i} (s_y r_z - s_z r_y, s_z r_x - s_x r_z, s_x r_y - s_y r_x) \cdot \boldsymbol{\sigma}.
%\end{align*}
%using $[\sigma_x,\sigma_y] = \mathrm{i} \sigma_z$ etc. Thus, $\mathrm{ad}_{\boldsymbol{s}\cdot
%\boldsymbol{\sigma}}$ can be represented by the matrix $\mathrm{i} S$ where
%\[  S = \left(
%      \begin{array}{ccc}
%        0 & -s_z & s_y \\
%        s_z & 0 & -s_x \\
%        -s_y & s_x & 0 \\
%      \end{array}
%    \right),
%\]
Note that the powers of the $\mathrm{ad}$ operator can be written in terms of the matrix $\boldsymbol{S}$ using the relation (\ref{eq:adhat}),
\begin{linenomath*}
\[ \mathrm{ad}_{-\mathrm{i}
\boldsymbol{s}_n\cdot \boldsymbol{\sigma}}(\widetilde{\boldsymbol{r}}) = \widetilde{\boldsymbol{S}_n \boldsymbol{r}} = - \mathrm{i} (\boldsymbol{S}_n \boldsymbol{r}) \cdot \boldsymbol{\sigma}, \]
\end{linenomath*}
and
\begin{linenomath*}
\[ \mathrm{ad}_{-\mathrm{i}
\boldsymbol{s}_n\cdot \boldsymbol{\sigma}}^p(\widetilde{\boldsymbol{r}}) = \widetilde{\boldsymbol{S}_n^p \boldsymbol{r}} =  - \mathrm{i} (\boldsymbol{S}_n^p \boldsymbol{r}) \cdot \boldsymbol{\sigma},
\quad p \geq 0.\]
\end{linenomath*}
Consequently,
\begin{linenomath*}
\begin{align*}
\frac{\partial \mathrm{U}_n}{\partial \theta_{n,k}}  &= - \mathrm{i} \mathrm{U}_n \left[\underbrace{\left(\sum_{p=0}^\infty \frac{(-\boldsymbol{S}_n)^p}{(p+1)!}
\right)}_{3\times 3}
\underbrace{\left( \frac{\partial \boldsymbol{s}_n }{\partial \theta_{n,k}} \right)}_{{3\times 1}} \right]\cdot \boldsymbol{\sigma},
\end{align*}
\end{linenomath*}
where $\boldsymbol{S}_n$ is obtained from $\boldsymbol{s}_n(\theta_{n,k})$ using equation~(\ref{eq:S}). Observe that
\begin{linenomath*}
\[ \boldsymbol{S}_n^3 = - \left\|\boldsymbol{s}_n\right\|^2 \boldsymbol{S}_n,\]
\end{linenomath*}
%so that
%\[ \boldsymbol{S}_n^{2k} = (-\norm{}{\boldsymbol{s}_n}^2)^{k-1} \boldsymbol{S}_n^2,\quad k \geq 1, \qquad \boldsymbol{S}_n^{2k+1} = (-\norm{}{\boldsymbol{s}_n}^2)^k \boldsymbol{S}_n,\quad k \geq 0,\]
and we may further simplify the dexp series as
\begin{linenomath*}
\begin{align}
\nonumber \boldsymbol{D}_n &= \sum_{p=0}^\infty \frac{(-\boldsymbol{S}_n)^p}{(p+1)!}\\&= I + \boldsymbol{S}_n^2 \sum_{p=1}^\infty \nonumber \frac{(-\left\|\boldsymbol{s}_n\right\|^2)^{p-1}}{(2p+1)!} - \boldsymbol{S}_n \sum_{p=0}^\infty \frac{(-\left\|\boldsymbol{s}_n\right\|^2)^p}{(2p+2)!}\\
\label{eq:D}&= I + c_1(\left\|\boldsymbol{s}_n\right\|) \boldsymbol{S}_n + c_2(\left\|\boldsymbol{s}_n\right\|) \boldsymbol{S}_n^2,
\end{align}
\end{linenomath*}
where
\begin{linenomath*}
\begin{equation}
\label{eq:c}
c_1(x) = \frac{\cos(x)-1}{x^2}, \qquad  c_2(x) =  \frac{x -\sin(x)}{x^3}.
\end{equation}
\end{linenomath*}

%\begin{align*}
%\boldsymbol{D}_n= \sum_{k=0}^\infty \frac{(-S)^k}{(k+1)!} &= \sum_{k=0}^\infty \frac{S^{2k}}{(2k+1)!} - \sum_{k=0}^\infty \frac{S^{2k+1}}{(2k+2)!}\\
%&= I + S^2 \sum_{k=1}^\infty \frac{(-\norm{}{\boldsymbol{s}}^2)^{k-1}}{(2k+1)!} - S \sum_{k=0}^\infty \frac{(-\norm{}{\boldsymbol{s}}^2)^k}{(2k+2)!}\\
%&= I + \left(\frac{\cos(\norm{}{\boldsymbol{s}})-1}{\norm{}{\boldsymbol{s}}^2}\right) S - \left(\frac{\sin(\norm{}{\boldsymbol{s}})-\norm{}{\boldsymbol{s}}}{\norm{}{\boldsymbol{s}}^3}\right) S^2,
%\end{align*}

To summarise, the analytic derivative of $\mathrm{U}_n$ is given by
\begin{equation}
\label{eq:dUn}
\frac{\partial \mathrm{U}_n}{\partial \theta_{n,k}} = - \mathrm{i} \mathrm{U}_n \left( \left[ \boldsymbol{D}_n \frac{\partial \boldsymbol{s}_n}{\partial \theta_{n,k}} \right] \cdot \boldsymbol{\sigma} \right),
\end{equation}
and the derivative of $\mathrm{U}_{\mathrm{tot}}$ by
\begin{linenomath*}
\begin{align}
\nonumber\frac{\partial \mathrm{U}_{\mathrm{tot}}}{\partial \theta_{n,k}} & = \mathrm{U}_{N} \mathrm{U}_{N-1} \ldots \mathrm{U}_{n+1} \frac{\partial \mathrm{U}_n}{\partial \theta_{n,k}} \mathrm{U}_{n-1} \ldots \mathrm{U}_{1}\\
\label{eq:gradUtot}& = - \mathrm{i} \mathrm{L}_{n} \left(\left[ \boldsymbol{D}_n \frac{\partial \boldsymbol{s}_n}{\partial \theta_{n,k}} \right] \cdot \boldsymbol{\sigma}\right) \mathrm{R}_{n-1}.
\end{align}
\end{linenomath*}
In a practical implementation, $\mathrm{L}_n$ and $\mathrm{L}_{n-1}$ are given by equations~(\ref{eq:L}) and (\ref{eq:R}). We compute $\boldsymbol{D}_n$ using equation~(\ref{eq:D}). Lastly, recall that in equation~(\ref{eq:s}),
\begin{linenomath*}
\[ \boldsymbol{s}_n =
(\Delta t) \, (f(t_{n-1}),g(t_{n-1}), \Omega)^\top,\]
\end{linenomath*}
and the control parameters are $\theta_{n,1} = f(t_{n-1})$ and $\theta_{n,2} = g(t_{n-1})$. Consequently,
\begin{equation}
\label{eq:ds1}
\frac{\partial \boldsymbol{s}_n}{\partial \theta_{n,1}} = (\Delta t)\, (1,0,0)^\top,
\end{equation}
and
\begin{equation}
\label{eq:ds2}
\frac{\partial \boldsymbol{s}_n}{\partial \theta_{n,2}} =  (\Delta t)\, (0,1,0)^\top.
\end{equation}
This completes the description of the analytic gradients.
Combining equations~(\ref{eq:gradF}), (\ref{eq:gradrhoN}) and (\ref{eq:gradUtot}),
\begin{equation}
\label{eq:gradFconcrete}
\frac{\partial
\mathcal{F}}{\partial \theta_{n,k}} = 2 \mathrm{Im} \mathrm{Tr} \left( \varrho^{\dagger} \mathrm{L}_{n} \left(\left[ \boldsymbol{D}_n \frac{\partial \boldsymbol{s}_n}{\partial \theta_{n,k}} \right] \cdot \boldsymbol{\sigma}\right) \mathrm{R}_{n-1}  \rho_0 \mathrm{U}_{\mathrm{tot}}^{\dagger}\right).
\end{equation}
Using the definition of $\mathrm{L}_n$ and $\mathrm{R}_{n}$ in equations~(\ref{eq:L}) and (\ref{eq:R}) it is evident that $\mathrm{L}_n \mathrm{R}_{n-1}=\mathrm{U}_{\mathrm{tot}}$ and consequently we may write
\begin{equation}
\label{eq:LRUtot}
\mathrm{R}_{n-1}=\mathrm{L}_n^{\dagger} \mathrm{U}_{\mathrm{tot}}.
\end{equation}
Substituting (\ref{eq:LRUtot}) in (\ref{eq:gradFconcrete}) reduces the final form of the analytical gradient of the fidelity function to
\begin{equation}
\label{eq:gradFinal}
\frac{\partial
\mathcal{F}}{\partial \theta_{n,k}} = 2 \mathrm{Im} \mathrm{Tr}( \mathcal{L}_{n,k}  \rho_N \varrho^{\dagger} ).
\end{equation}
where for any pulse segment $n$ and any control parameter $k$:
\begin{equation}
\label{eq:LDnk}
\mathcal{L}_{n,k}  = \mathrm{L}_{n} \left(\left[ \boldsymbol{D}_n \frac{\partial \boldsymbol{s}_n}{\partial \theta_{n,k}} \right] \cdot \boldsymbol{\sigma}\right) \mathrm{L}_{n}^{\dagger},
\end{equation}

%
%Note that $\mathrm{U}_n$ is already computed and can be recombined with $\mathrm{L}$, as we
%will see shortly. Secondly, $D_n$ is independent of $\theta_{n,k}$ in the sense
%that we need to compute it once for every $\mathrm{U}_n$ (i.e. $\boldsymbol{s}_n$,
%effectively), but not once for every $\theta_{n,k}$. Only $\frac{\partial
%\boldsymbol{s}_n}{\partial \theta_{n,k}}$ needs to be computed for every $\theta_{n,k}$,
%which is assumed to be inexpensive. For instance, in piecewise constant GRAPE
%type approach, where $\theta_{n,k}$ might be $f_{j-1} = f(t_{j-1})$, the
%derivative $\frac{\partial \boldsymbol{s}_n}{\partial \theta_{n,k}} = \delta_{n,k}
%(\Delta t) (1,0,0)$.
%
%
%Later a comment that $\frac{\partial \boldsymbol{s}_n(\theta_{n,k}) }{\partial \theta_{n,k}}$
%is available.

%\begin{align*}
%\frac{\partial \Utot}{\partial \theta_{n,k}} & = \sum_{n=1}^N \mathrm{U}_{N} \mathrm{U}_{N-1} \ldots \mathrm{U}_{n+1} \frac{\partial \mathrm{U}_n}{\partial \theta_{n,k}} \mathrm{U}_{n-1} \ldots \mathrm{U}_{1}\\
%& = \sum_{n=1}^N \mathrm{L}_{n+1} \frac{\partial \mathrm{U}_n}{\partial \theta_{n,k}} \mathrm{R}_{n-1}\\
%& = - \mathrm{i}\sum_{n=1}^N \mathrm{L}_{n+1} \mathrm{U}_n \left(\left[ D_n \left( \frac{\partial \boldsymbol{s}_n(\theta_{n,k}) }{\partial
%\theta_{n,k}} \right) \right] \cdot \boldsymbol{\sigma}\right) \mathrm{R}_{n-1}\\
%& = - \mathrm{i}\sum_{n=1}^N \mathrm{L}_{n} \left(\left[ D_n \left( \frac{\partial \boldsymbol{s}_n(\theta_{n,k}) }{\partial
%\theta_{n,k}} \right) \right] \cdot \boldsymbol{\sigma}\right) \mathrm{R}_{n-1}.
%\end{align*}

\label{subsec:grad}

\subsection{Computation of Hessian}
In the computation of the Hessian of the fidelity function (\ref{eq:hessF}) we require the Hessian of the final state,
\begin{align}
\nonumber
 \frac{\partial^2 \rho_N}{\partial \theta_{m,j}\partial \theta_{n,k} }   &= 2 \mathrm{Re} \left( \frac{\partial^2 \mathrm{U}_{\mathrm{tot}}}{ \partial \theta_{m,j} \partial \theta_{n,k}}  \rho_0 \mathrm{U}_{\mathrm{tot}}^{\dagger} \right.\\
\label{eq:hessrhoN} & \qquad \qquad + \left.
 \frac{\partial \mathrm{U}_{\mathrm{tot}}}{\partial \theta_{n,k}}  \rho_0 \frac{\partial \mathrm{U}_{\mathrm{tot}}}{\partial \theta_{m,j}}^{\dagger} \right) .
 \end{align}
An analytic form for the gradient of $\mathrm{U}_{\mathrm{tot}}$ with respect to control parameters $\theta_{n,k}$ and $\theta_{m,j}$ has already been obtained in equation~(\ref{eq:gradUtot}). In this section, we derive an analytic form for $\partial^2 \mathrm{U}_{\mathrm{tot}}/\partial \theta_{m,j}\partial \theta_{n,k} $.

\subsubsection{Off-diagonal entries ($n>m$) of the Hessian}
When $n>m$, the Hessian is typically computed as
\begin{align}
\nonumber \frac{\partial^2 \mathrm{U}_{\mathrm{tot}}}{ \partial \theta_{m,j} \partial \theta_{n,k}}  &= \underbrace{\mathrm{U}_{N} \mathrm{U}_{N-1} \ldots \mathrm{U}_{n+1}}_{\mathrm{L}_{n+1}} \frac{\partial \mathrm{U}_{n}}{\partial \theta_{n,k}}  \\
\nonumber & \qquad \times
\underbrace{\mathrm{U}_{n-1} \ldots \mathrm{U}_{m+1}}_{\mathrm{M}_{n-1,m+1}} \frac{\partial \mathrm{U}_{m}}{\partial \theta_{m,j}} \\
\label{eq:hessUtotmn} & \qquad \quad \times \underbrace{\mathrm{U}_{m-1} \ldots \mathrm{U}_{1}}_{\mathrm{R}_{m-1}}.
\end{align}
where
\begin{align}
\mathrm{M}_{n,m} & = \mathrm{U}_{n} \mathrm{U}_{n-1} \ldots  \mathrm{U}_{m-1} \mathrm{U}_{m}.
\end{align}
Similarly, we can derive the corresponding expression for $m>n$. Overall, since $n$ and $m$ range between $1$ and $N$, the various values of $\mathrm{M}_{n,m}$ are typically computed in $\mathcal{O}(N^2)$ time in such a procedure.

Here we introduce an alternative approach for computing $\partial^2 \mathrm{U}_{\mathrm{tot}}/ \partial \theta_{m,j} \partial \theta_{n,k}$ that does not require the computation of $\mathrm{M}_{n,m}$.
Since $\mathrm{L}_{n+1}$ and $\mathrm{R}_{m-1}$ are  unitary,
\begin{equation}
\mathrm{L}_{n+1}^{\dagger}\mathrm{L}_{n+1}^{}=I, \qquad \mathrm{R}_{m-1}^{\dagger}\mathrm{R}_{m-1}^{}=I,
\end{equation}
we can express $\mathrm{M}_{n,m}$ as
\begin{align}
\nonumber \mathrm{M}_{n,m} & = \mathrm{U}_{n} \mathrm{U}_{n-1} \ldots \mathrm{U}_{m}\\
\nonumber & = \mathrm{L}_{n+1}^{\dagger}\mathrm{L}_{n+1} \mathrm{U}_{n} \mathrm{U}_{n-1} \ldots \mathrm{U}_{m} \mathrm{R}_{m-1}\mathrm{R}_{m-1}^{\dagger}\\
\label{eq:Mid}& = \mathrm{L}_{n+1}^{\dagger} \mathrm{U}_{\mathrm{tot}} \mathrm{R}_{m-1}^{\dagger}.
\end{align}

Thus, $\mathrm{M}_{n,m}$ can be replaced in the computation of the Hessian and equation~(\ref{eq:hessUtotmn}) can be written in the form
\begin{equation}
\label{eq:hessUtotmn1}
\frac{\partial^2 \mathrm{U}_{\mathrm{tot}}}{\partial \theta_{m,j} \partial \theta_{n,k} }  = \mathrm{L}_{n+1} \frac{\partial \mathrm{U}_{n}}{\partial \theta_{n,k}}  \mathrm{L}_{n}^{\dagger} \mathrm{U}_{\mathrm{tot}} \mathrm{R}_{m}^{\dagger}   \frac{\partial \mathrm{U}_{m}}{\partial \theta_{m,j}} \mathrm{R}_{m-1}.
\end{equation}
Substituting (\ref{eq:dUn}), the expression in equation~(\ref{eq:hessUtotmn1}) becomes
\begin{align*}
&-  \mathrm{L}_{n+1} \mathrm{U}_n  \left( \left[ \boldsymbol{D}_n \frac{\partial \boldsymbol{s}_n}{\partial \theta_{n,k}} \right] \cdot \boldsymbol{\sigma}\right) \mathrm{L}_{n}^{\dagger} \mathrm{U}_{\mathrm{tot}}  \\
& \qquad \qquad \quad \times \mathrm{R}_{m}^{\dagger} \mathrm{U}_m \left( \left[ \boldsymbol{D}_m \frac{\partial \boldsymbol{s}_m}{\partial \theta_{m,j}} \right] \cdot \boldsymbol{\sigma}\right) \mathrm{R}_{m-1}.
\end{align*}
We use the fact that $\mathrm{L}_{n+1} \mathrm{U}_n = \mathrm{L}_n$ and $\mathrm{R}_{m}^{\dagger} \mathrm{U}_m = \mathrm{R}_{m-1}^{\dagger}$ to reduce this expression to
\begin{align*}
\label{eq:hessUtotLR}
&- \mathrm{L}_{n} \left( \left[ \boldsymbol{D}_n \frac{\partial \boldsymbol{s}_n}{\partial \theta_{n,k}} \right] \cdot \boldsymbol{\sigma}\right) \mathrm{L}_{n}^{\dagger} \mathrm{U}_{\mathrm{tot}} \\
& \qquad \qquad \quad \times \mathrm{R}_{m-1}^{\dagger}  \left( \left[ \boldsymbol{D}_m \frac{\partial \boldsymbol{s}_m}{\partial \theta_{m,j}} \right] \cdot \boldsymbol{\sigma}\right) \mathrm{R}_{m-1}.
\end{align*}

%where
%\begin{align}
%\mathcal{L}_{n,k} &=  \mathrm{L}_n  \left( \left[ \boldsymbol{D}_n \frac{\partial \boldsymbol{s}_n}{\partial \theta_{n,k}} \right] \cdot \boldsymbol{\sigma}\right) \adj{\mathrm{L}_{n}} \\
%\RD_{m,j} & = \adj{\mathrm{R}_{m-1}}\left( \left[ \boldsymbol{D}_m \frac{\partial \boldsymbol{s}_m}{\partial \theta_{m,j}} \right] \cdot \boldsymbol{\sigma}\right) \mathrm{R}_{m-1}.
%\end{align}

\subsubsection{Diagonal entries ($m=n$) of the Hessian}
For the case $m=n$, following (\ref{eq:gradLR}),
\begin{equation}
\label{eq:hessUtotnn}
\frac{\partial^2 \mathrm{U}_{\mathrm{tot}}}{\partial \theta_{n,j} \partial \theta_{n,k}}  = \mathrm{L}_{n+1} \frac{\partial^2 \mathrm{U}_n}{\partial \theta_{n,j} \partial \theta_{n,k}}
\mathrm{R}_{n-1}.
\end{equation}

Differentiating equation~(\ref{eq:dUn}) with respect to $\theta_{n,j}$,
\begin{align}
\nonumber & \frac{\partial^2 \mathrm{U}_n}{\partial \theta_{n,j} \partial \theta_{n,k}} = \\
\nonumber & \qquad - \mathrm{U}_n \left\{\left[ \left(\boldsymbol{D}_n \frac{\partial \boldsymbol{s}_n}{\partial \theta_{n,j}} \right) \cdot \boldsymbol{\sigma}\right]
\left[ \left(\boldsymbol{D}_n \frac{\partial \boldsymbol{s}_n}{\partial \theta_{n,k}} \right) \cdot \boldsymbol{\sigma}\right] \right.\\
\label{eq:hessUnn} &\qquad \left. +\mathrm{i} \left( \frac{\partial
\boldsymbol{D}_n}{\partial \theta_{n,j}} \frac{\partial \boldsymbol{s}_n}{\partial \theta_{n,k}} + \boldsymbol{D}_n \frac{\partial^2 \boldsymbol{s}_n}{\partial \theta_{n,j} \partial \theta_{n,k}} \right)  \cdot \boldsymbol{\sigma}
\right\},
\end{align}
where $\partial^2 \boldsymbol{s}_n/\partial \theta_{n,j} \partial \theta_{n,k}$ vanishes due to (\ref{eq:ds1}) and (\ref{eq:ds2}).
The derivative of $\boldsymbol{D}_n$ (\ref{eq:D}) can be computed explicitly,
\begin{align}
\frac{\partial \boldsymbol{D}_n}{\partial \theta_{n,j}} &= \frac{\partial}{\partial \theta_{n,j}} \left( I + c_1(\left\|\boldsymbol{s}_n\right\|) \boldsymbol{S}_n + c_2(\left\|\boldsymbol{s}_n\right\|) \boldsymbol{S}_n^2\right)\\
\nonumber & = c_1'(\left\|\boldsymbol{s}_n\right\|) \frac{\partial \left\|\boldsymbol{s}_n\right\|}{\partial \theta_{n,j}} \boldsymbol{S}_n + c_1(\left\|\boldsymbol{s}_n\right\|) \frac{\partial \boldsymbol{S}_n}{\partial \theta_{n,j}} \\
\nonumber & \qquad + c_2'(\left\|\boldsymbol{s}_n\right\|) \frac{\partial \left\|\boldsymbol{s}_n\right\|}{\partial \theta_{n,j}} \boldsymbol{S}_n^2 \\
\nonumber & \qquad+ c_2(\left\|\boldsymbol{s}_n\right\|) (\boldsymbol{S}_n \frac{\partial \boldsymbol{S}_n}{\partial \theta_{n,j}} + \frac{\partial \boldsymbol{S}_n}{\partial \theta_{n,j}} \boldsymbol{S}_n),
\end{align}
where
\begin{align}
c_1'(x) &= \frac{-2 \cos(x) - x \sin(x)  + 2 }{x^3},\\
c_2'(x) &= \frac{3 \sin(x) - x \cos(x) - 2x}{x^4} ,
\end{align}
\begin{equation}
\frac{\partial \left\|\boldsymbol{s}_n\right\|}{\partial \theta_{n,j}} = \frac{ \boldsymbol{s}_n \cdot  \frac{\partial \boldsymbol{s}_n}{\partial \theta_{n,j}}  }{\left\|\boldsymbol{s}_n\right\|},\end{equation}
and $\partial \boldsymbol{S}_n/\partial \theta_{n,j}$ is obtained directly by creating a matrix from $\partial \boldsymbol{s}_n/\partial \theta_{n,j}$, equations~(\ref{eq:ds1}) and (\ref{eq:ds2}), along the lines of (\ref{eq:S}),
\begin{equation}
\frac{\partial \boldsymbol{S}_n}{\partial \theta_{n,1}} = (\Delta t)   \left(
      \begin{array}{ccc}
        0 & 0 & 0 \\
        0 & 0 & -1 \\
        0 & 1 & 0 \\
      \end{array}
    \right), \ \frac{\partial \boldsymbol{S}_n}{\partial \theta_{n,2}} = (\Delta t)   \left(
      \begin{array}{ccc}
        0 & 0 & 1 \\
        0 & 0 & 0 \\
        -1 & 0 & 0 \\
      \end{array}
    \right).
\end{equation}

%\subsection{Hessian}

The complete description of the Hessian of the fidelity is obtained by combining equations~(\ref{eq:hessF}), (\ref{eq:hessrhoN}), (\ref{eq:hessUtotmn}), (\ref{eq:hessUnn}), (\ref{eq:hessUtotnn}) and (\ref{eq:gradUtot}),
\begin{equation}
\label{eq:hessFconcretemn}
\frac{\partial^2 \mathcal{F}}{\partial \theta_{m,j} \partial \theta_{n,k}} =  2 \mathrm{Re} \mathrm{Tr} (\varrho^{\dagger} (\mathcal{V}_{m,n,j,k} - \mathcal{W}_{m,n,j,k})),
\end{equation}
where
\begin{align}
\nonumber \mathcal{V}_{m,n,j,k}  &= \mathrm{L}_{n} \left(\left[ \boldsymbol{D}_n \frac{\partial \boldsymbol{s}_n}{\partial \theta_{n,k}} \right] \cdot \boldsymbol{\sigma}\right) \mathrm{R}_{n-1} \rho_0 \\
& \qquad \times {\left(\mathrm{L}_{m} \left( \left[ \boldsymbol{D}_m \frac{\partial \boldsymbol{s}_m}{\partial \theta_{m,j}} \right] \cdot \boldsymbol{\sigma}\right) \mathrm{R}_{m-1}   \right)}^{\dagger},
\end{align}
Using equations~(\ref{eq:LRUtot}) and (\ref{eq:LDnk}) we can simplify the above expressions as follows:
\begin{equation}
\label{eq:Vreduced}
\mathcal{V}_{m,n,j,k}  = \mathcal{L}_{n,k} \rho_N \mathcal{L}_{m,j}^{\dagger}
\end{equation}
%where for any pulse segment $n$ and and control parameter $k$:
%
%\begin{equation}
%\label{eq:LDnk}
%\mathcal{L}_{n,k}  = \mathrm{L}_{n} \left(\left[ \boldsymbol{D}_n \frac{\partial \boldsymbol{s}_n}{\partial \theta_{n,k}} \right] \cdot \boldsymbol{\sigma}\right) \adj{\mathrm{L}_{n}},
%\end{equation}
while $\mathcal{V}_{m,n,j,k}$ is common between diagonal ($m=n$) and off-diagonal elements ($n>m$) of the Hessian matrix $\mathcal{W}_{m,n,j,k}$ has two distinct forms.\\
For $m=n$,
\footnotesize
\begin{align*}
\mathcal{W}_{n,n,j,k} &=  \mathrm{L}_{n}  \left\{\left[ \left(\boldsymbol{D}_n \frac{\partial \boldsymbol{s}_n}{\partial \theta_{n,j}} \right) \cdot \boldsymbol{\sigma}\right]
\left[ \left(\boldsymbol{D}_n \frac{\partial \boldsymbol{s}_n}{\partial \theta_{n,k}} \right) \cdot \boldsymbol{\sigma}\right] \right. \\
& \qquad \quad \left.+\mathrm{i} \left[ \left( \frac{\partial
\boldsymbol{D}_n}{\partial \theta_{n,j}} \frac{\partial \boldsymbol{s}_n}{\partial \theta_{n,k}} \right)  \cdot \boldsymbol{\sigma} \right]
\right\} \mathrm{R}_{n-1} \rho_0 \mathrm{U}_{\mathrm{tot}}^{\dagger},
\end{align*}
\normalsize
and for $n>m$,
\begin{align*}
%\label{eq:hessFconcretemn}
\mathcal{W}_{m,n,j,k} &= \mathrm{L}_n  \left( \left[ \boldsymbol{D}_n \frac{\partial \boldsymbol{s}_n}{\partial \theta_{n,k}} \right] \cdot \boldsymbol{\sigma}\right) \mathrm{L}_{n}^{\dagger}  \mathrm{U}_{\mathrm{tot}} \\
& \qquad \times \mathrm{R}_{m-1}^{\dagger}\left( \left[ \boldsymbol{D}_m \frac{\partial \boldsymbol{s}_m}{\partial \theta_{m,j}} \right] \cdot \boldsymbol{\sigma}\right) \mathrm{R}_{m-1} \rho_0 \mathrm{U}_{\mathrm{tot}}^{\dagger}.
\end{align*}

Similar to equation~(\ref{eq:Vreduced})  we can do additional simplifications for the $\mathcal{W}_{m,n,j,k}$ term.
For $m=n$ we have:
\begin{equation}
\mathcal{W}_{n,n,j,k}  = \left(\mathcal{L}_{n,j}  \mathcal{L}_{n,k} + \mathcal{D}_{n,j,k}\right) \rho_N
\end{equation}
where
\begin{equation}
\mathcal{D}_{n,j,k} =  \mathrm{i} \mathrm{L}_{n}  \left[\left( \frac{\partial
\boldsymbol{D}_n}{\partial \theta_{n,j}} \frac{\partial \boldsymbol{s}_n}{\partial \theta_{n,k}} \right)  \cdot \boldsymbol{\sigma} \right] \mathrm{L}_{n}^{\dagger}.
\end{equation}
and for $n>m$ we can use equation~(\ref{eq:LDnk}) to write:
\begin{equation}
\mathcal{W}_{m,n,j,k}  = \mathcal{L}_{n,k} \mathcal{L}_{m,j} \rho_N
\end{equation}
Therefore the general form of the diagonal elements of the Hessian matrix will be:
\begin{align}
\nonumber \frac{\partial^2 \mathcal{F}}{\partial \theta_{n,j} \partial \theta_{n,k}} &=  2 \mathrm{Re} \mathrm{Tr} (  \mathcal{L}_{n,k} \rho_N \mathcal{L}_{n,j}^{\dagger} \varrho^{\dagger}\\
\label{eq:hessFinalDiag}& \qquad \qquad -(\mathcal{L}_{n,j}  \mathcal{L}_{n,k} + \mathcal{D}_{n,j,k}) \rho_N \varrho^{\dagger}),
\end{align}
and the general form of the upper-diagonal elements of the Hessian matrix can be written as
\begin{equation}
\label{eq:hessFinaloffDiag}
\frac{\partial^2 \mathcal{F}}{\partial \theta_{m,j} \partial \theta_{n,k}} =  2 \mathrm{Re} \mathrm{Tr} (\mathcal{L}_{n,k} \rho_N \mathcal{L}_{m,j}^{\dagger} \varrho^{\dagger}  -\mathcal{L}_{n,k} \mathcal{L}_{m,j} \rho_N \varrho^{\dagger}),
\end{equation}
$\mathcal{L}$ and $\mathcal{D}$ can be precomputed in $\mathcal{O}(N)$ time along with $\mathrm{L}$. The factorization (\ref{eq:LDnk}) reduces the computational effort by a factor of three since only two matrix multiplications are required for each entry of the Hessian.
Note that the lower-diagonal elements ($n<m$) can be easily obtained using the symmetry of the Hessian matrix and do not need to be computed separately. Equation (\ref{eq:hessFinaloffDiag}) for these entries can be written as:
\begin{equation}
\label{eq:hessFinaloffDiag2}
\frac{\partial^2 \mathcal{F}}{\partial \theta_{m,j} \partial \theta_{n,k}} =  2 \mathrm{Re} \mathrm{Tr} (\mathcal{L}_{n,k} \rho_N \mathcal{L}_{m,j}^{\dagger}\varrho^{\dagger}  -\mathcal{L}_{m,j} \mathcal{L}_{n,k} \rho_N \varrho^{\dagger}),
\end{equation}
Finally, using equations (\ref{eq:hessFinalDiag}), (\ref{eq:hessFinaloffDiag}) and (\ref{eq:hessFinaloffDiag2}) a general form of Hessian entries can be expressed as a single equation:
\begin{align}
\nonumber \frac{\partial^2 \mathcal{F}}{\partial \theta_{m,j} \partial \theta_{n,k}} &=  2 \mathrm{Re} \mathrm{Tr} (\mathcal{L}_{n,k} \rho_N \mathcal{L}_{m,j}^{\dagger} \varrho^{\dagger} \\
\label{eq:hessFinalFinal} & \qquad  -(\underbrace{\mathcal{L}_{n,k}  \mathcal{L}_{m,j}}_{\underset{\text{if }m > n}{\mathcal{L}_{m,j}  \mathcal{L}_{n,k}}} + \underbrace{\mathcal{D}_{n,j,k}}_{\underset{\text{if }n \ne m}{0}}) \rho_N \varrho^{\dagger} ),
\end{align}

\label{subsec:hess}

\section{Numerical demonstrations}
\subsection{Comparison with finite difference method}
\begin{figure}
  \includegraphics[width=\linewidth]{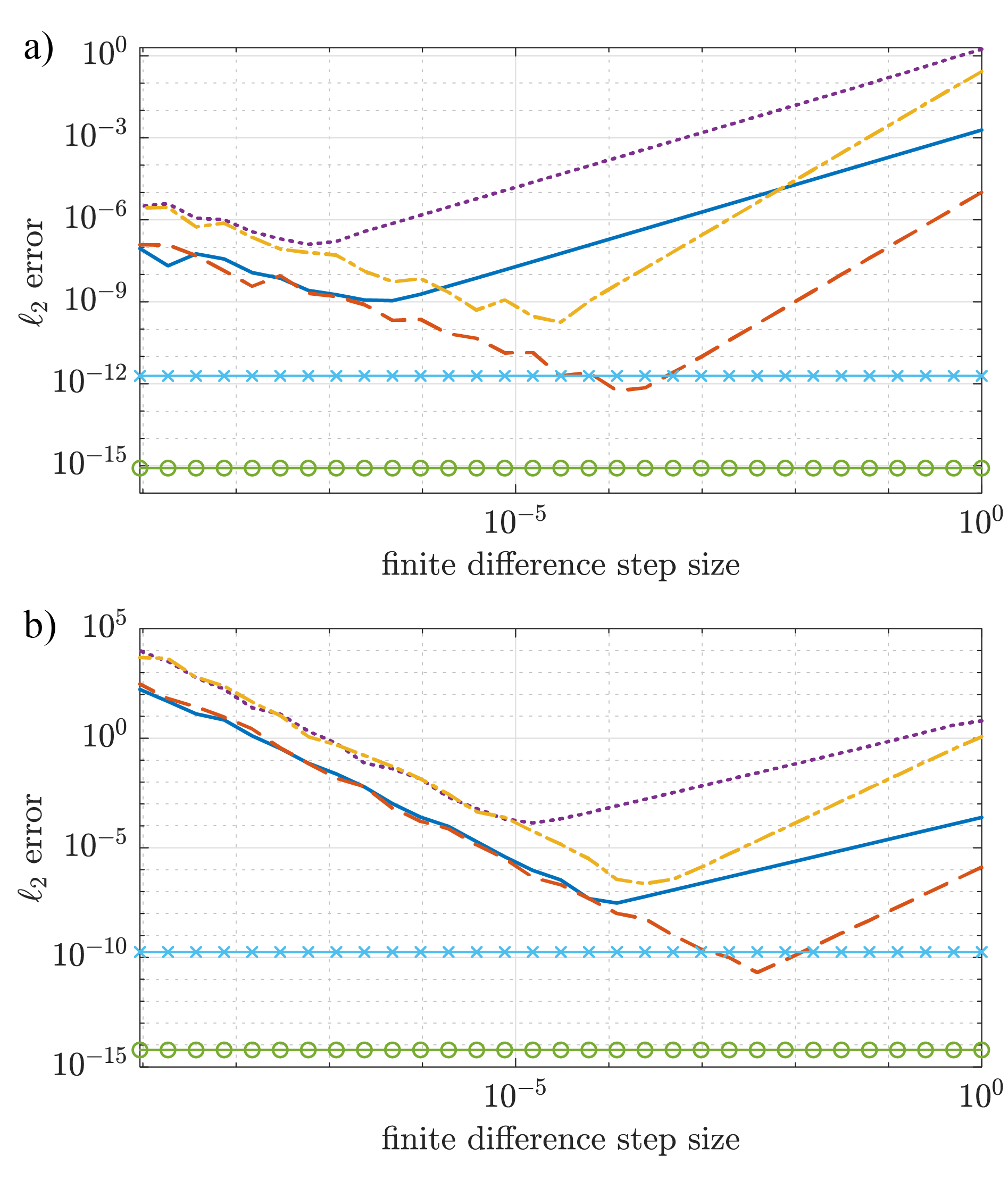}\\
  \caption{Accuracy and stability of forward differences and central differences approximation to (a) the gradient ($\partial U_n/\partial \theta_{n,k}$), and (b) the Hessian ($\partial^2 U_n/\partial \theta_{n,k}^2$) of a single step propagator $U_n$. The behaviour is shown for two different time steps, $\Delta t = 10^{-6}$ (forward difference [solid blue], central difference [dotted purple], analytic [solid green with circles]) and $\Delta t = 10^{-4}$ (forward difference [dashed orange], central difference [densely dashed yellow], analytic [solid light blue with crosses]).}
  \label{fig:Fig_1}
\end{figure}

Finite difference approximation of the gradient of numerical propagators $U_n$ requires computing $U_n$ for multiple values of $\theta_{n,k}$ differing by a `{\em finite difference step}'. Figure~\ref{fig:Fig_1} demonstrates that while the finite difference step size must be kept sufficiently small for accuracy, the approximations become unstable for very small steps. Thus the suitability of a finite difference step may prove difficult to asses a-priori. This balance between accuracy and stability becomes more precarious when (i) the time step of the numerical propagator ($\Delta t$) is large, (ii) the numerical propagator is of limited accuracy or (iii) higher derivatives are required. In addition to respecting the Lie algebraic structure and being relatively inexpensive, the proposed approach for computing analytic derivatives does not suffer from such instability.

\subsection{Example for pulse design in magnetic resonance}

Here we demonstrate one of the applications of the proposed method for the design of broadband excitation pulses in magnetic resonance spectroscopy. The simplest case would be control of an ensemble of non-interacting spin-\sfrac{1}{2} particles. Conventional instantaneous radio-frequency or microwave pulses have limited bandwidth due to high power requirements that cannot be afforded on most instruments; therefore they can only satisfy the desired state manipulation in a rather limited range of frequencies close to the transmitter offset of the pulse, i.e. they are only effective for spins with relatively small frequency offsets; additionally, the performance of these pulses can be considerably affected by instrumental imperfections or instabilities. The goal here is to circumvent these problems by designing a pulse propagator that satisfies certain objectives for all spins within the desired frequency range, with a robust performance that does not depend on frequency offset of spins or instrumental imperfections.

\begin{figure}
  \includegraphics[width=\linewidth]{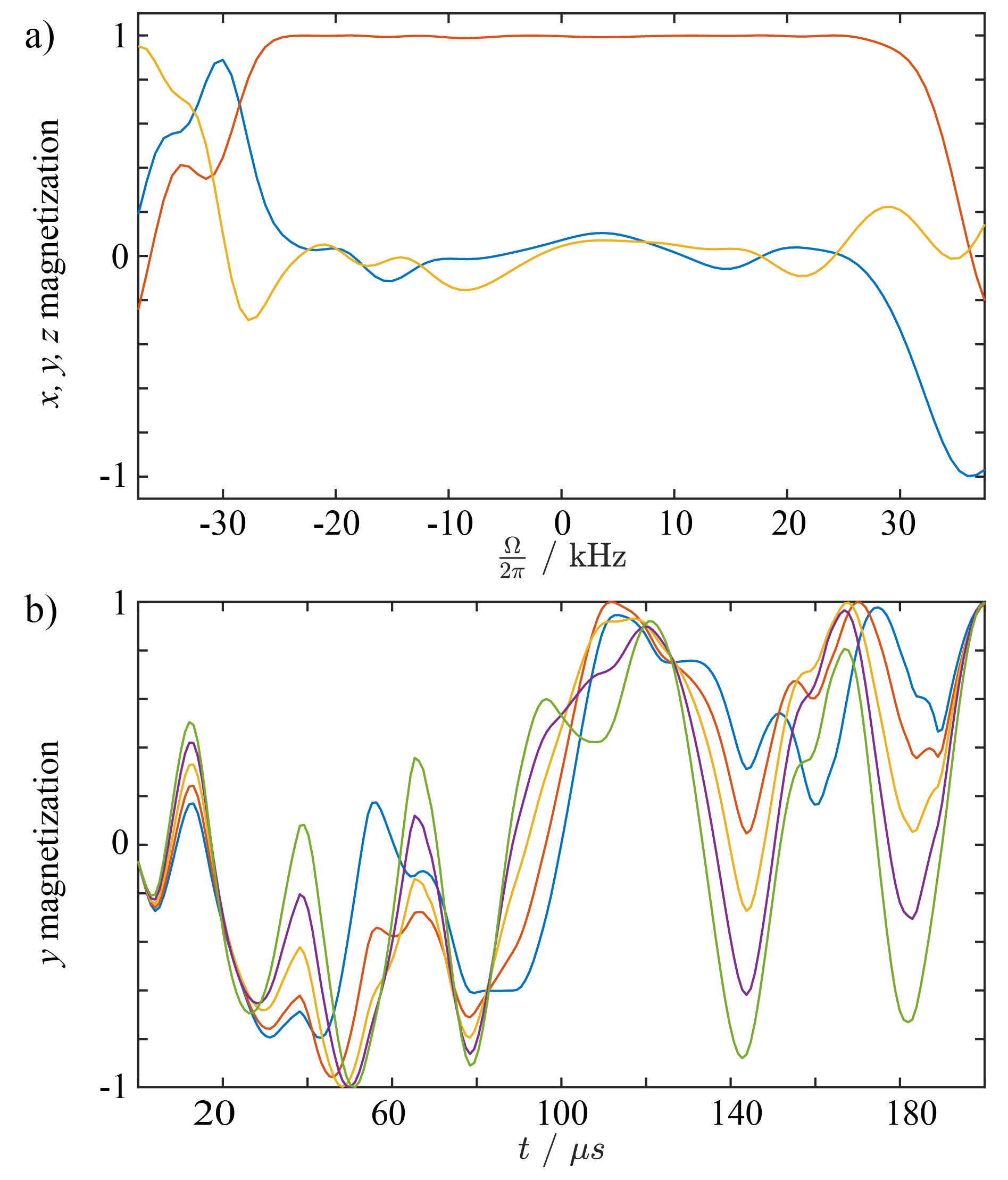}
  \caption{(a) calculated excitation profiles for different component of density operator, $x$ (blue), $y$ (red), and $z$ (orange) at $\omega_{1}=\omega_{1}^0$ (20 kHz) for a 200 $\mu$s pulse acting on 101 non-interacting spin-\sfrac{1}{2} over a 50 kHz frequency range; (b) variations of $y$ component of spin trajectories during the pulse for five different frequencies.}
  \label{fig:Fig_2}
\end{figure}

\begin{figure}
  \includegraphics[width=\linewidth]{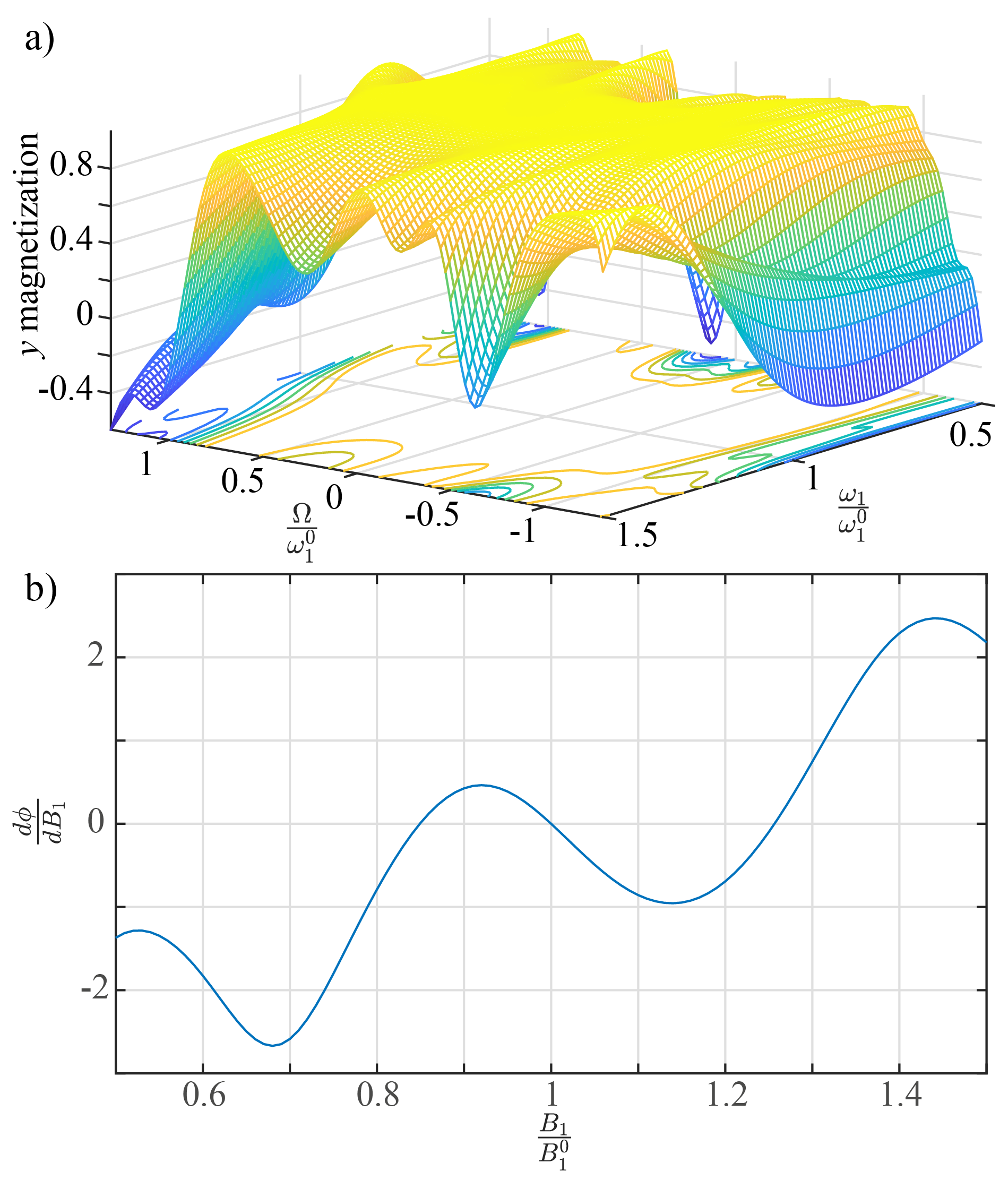}
  \caption{(a) 3D plot and projection showing the $y$-magnetization excited as a function of relative resonance offset ($\frac{\Omega}{\omega_{1}^0}$) and relative RF amplitude $\frac{\omega_{1}}{\omega_{1}^0}$; (b) Sensitivity of signal phase to field strength  ($\frac{\mathrm{d}\phi}{\mathrm{d}B_{1}}$) as a function of relative field strength ($\frac{B_{1}}{B_{1}^0}$).}
  \label{fig:Fig_3}
\end{figure}

The example here demonstrates an excitation pulse designed using the proposed method to bring all spins in the ensemble from $z$ to $y$. Figure~\ref{fig:Fig_2} (a) shows the final state of spin across the frequency range of interest (50 kHz here), and figure~\ref{fig:Fig_2} (a) shows variations of one of the components, $y$, for five different offset frequencies during the 200 $\mu$s pulse. Additionally, we can incorporate an additional optimization step that significantly reduces the sensitivity of the pulse to instrumental imperfections. Here this was considered as reducing the sensitivity of the pulse performance to unknown variations of radio-frequency (RF), or microwave (MW) amplitudes. Figure~\ref{fig:Fig_3} (a) shows corresponding graphs for the variations of the target state, $y$, versus RF field, $B_{1}$. One common example of such imperfection is the position-dependent $B_{1}$ field across an RF coil used to generate the pulse. These variations introduce position-dependent phase of the signal across the ensemble of spins and therefore results in significant signal loss and non-uniform excitation profile of the pulse. Here an additional objective is to minimize the variation of signal phase with respect to the variation of $B_{1}$ field ($\frac{\mathrm{d}\phi}{\mathrm{d}B_{1}}$), figure~\ref{fig:Fig_3} (b) shows that for a given nominal RF amplitude with $\pm 20\%$ variations in the amplitude of $B_{1}$ field, $\frac{\mathrm{d}\phi}{\mathrm{d}B_{1}}$ is zero for all frequencies in the desired range.

\label{sec:num}

\section{Conclusion}

In the present work, we have introduced a new approach, ESCALADE, for computation of derivatives of the cost function in optimal control of spin systems. We demonstrated that using the proposed mathematical framework, derivatives (gradient and Hessian) can be computed analytically using Lie algebraic techniques. The proposed method is very general and can be adapted to and used in many potential applications where efficient optimal control of spin systems is required. A numerical implementation of the proposed method in MATLAB along with additional functions for optimization and visualization of the performance are freely available via the following DOI: 10.17632/8zz84359m5.1.

\section*{Acknowledgements}
MF thanks the Royal Society for a University Research Fellowship and a University Research Fellow Enhancement Award  (grant numbers URF\textbackslash R1\textbackslash180233 and RGF\textbackslash EA\textbackslash181018).
PS thanks Trinity College Oxford for a Junior Research Fellowship and Mathematical Institute, Oxford, where most of this research was carried out. 

\biboptions{numbers,sort&compress}
\bibliographystyle{elsarticle-num}       
\bibliography{ms}           

\end{document}